\documentclass[10pt,twoside]{hsqcd}
\usepackage[dvips]{graphicx,epsfig,color}
\usepackage{wrapfig,rotating}
\usepackage{amssymb,amsmath,array}
\input{symblibrary.tex}
\begin{document}
\title{BEYOND THE STANDARD MODEL \\
       PHYSICS SEARCHES AT THE TEVATRON 
\thanks{The talk given at the workshop 
        ``Hadron Structure and QCD: from LOW to HIGH energies'', 
         HSQCD 2008, Gatchina, Russia, June~30~-~July~4,~2008
       }}
\author{Igor~V.~Gorelov \\ \\
{\it for the CDF and \( D\O \) Collaborations} \\ \\
    Department of Physics and Astronomy, \\
    MSC07 4220, University of New Mexico, \\
    800 Yale Blvd. NE, Albuquerque, NM 87131, USA \\
    E-mail: gorelov@fnal.gov }
\maketitle
\begin{abstract} 
\noindent 
  The recent results on a number of searches performed at Tevatron for
  new phenomena beyond Standard Model are presented. The topics
  include the experimental tests of SUSY with mSUGRA and GMSB breaking
  scenarios. The latest analyses on large extra dimensions, new
  massive bosons are covered as well. The results are based on
  experimental data samples collected at the Tevatron with CDF and \dzero
  detectors and comprising a total integrated luminosity up to 
  \( \sim2.7\invfb \).
\end{abstract}
%
%
%
  \markboth{\large \sl {Igor~V.~Gorelov} 
  \hspace*{2cm} HSQCD 2008} {\large \sl \hspace*{1cm} 
   Beyond the Standard Model Physics Searches at the Tevatron }
%
\section{Experimental Apparatus} 
  CDF and \dzero Collaborations are running experiments at the Tevatron
  collider collecting data from \( \pap \) collisions at an energy of
  \( \sqs = 1960\gev \). The \cdf2 (later CDF) and \dzero are upgraded
  multipurpose high-energy physics detectors~\cite{Acosta:2004yw}
  with silicon vertex detectors, central
  tracking, electromagnetic and hadron calorimetry and muon
  identification systems. To the date of this presentation the
  Tevatron machine delivered more than \( 4\invfb \) integrated
  luminosity. The analyses to be discussed below are based on an
  amount of data corresponding to integrated luminosity 
  \(\IntL\) up to \( \sim2.7\invfb \).
\section{SUSY Searches at the Tevatron}
  Supersymmetry (SUSY) transforms fermions into bosons and vice versa.
  The SUSY theoretical approach offers solutions for several
  theoretical and experimental challenges of modern high-energy
  physics. It resolves the divergences inherent to a Standard Model
  (``hierarchy problem'') and  provides candidates for Dark Matter.
  It creates harmony with unification theories at the Planck scale and
  embodies gravity opening a path to a string theory.  Since
  superpartners of known Standard Model (SM) particles have not yet
  been observed, SUSY is a broken symmetry.  In our searches we
  consider a Minimal SUSY Model (MSSM) within its two breaking
  scenarios, mSUGRA and GMSB, as the benchmarks for the theoretical
  predictions.  mSUGRA is broken by gravity at the Planck scale. With
  R-parity invariance~\cite{def:Rp}, the mSUGRA spectrum and its
  interaction strengths are determined by 5 parameters: soft breaking
  scalar and fermion scales \( m_{0},m_{1/2}\), trilinear coupling
  \(A_{0}\), ratio \( \tan{\beta}=\langle\,H_{u}\rangle / \langle\,H_{d}\rangle \), 
  and sign of the trilinear coupling in a
  Higgsino mass term \( sgn(\mu_{0}) \).  The lightest SUSY particle
  (LSP) in mSUGRA is the neutralino (\(\snone \)).  In the MSSM with
  gauge-mediated supersymmetry-breaking (GMSB) scenario, the breaking
  is communicated via gauge fields (e.g. EWK or QCD) and at a much lower
  scale of \( \Lambda\sim100\tev \). The minimal GMSB scenario
  is again R-parity invariant and is determined by \(\Lambda \), mass
  of a messenger field \( M_{mess} \), number of messenger fields 
  \( N_{5} \), \(\tan{\beta} \) and \( sgn(\mu_{0}) \). The LSP in GMSB
  framework is the gravitino (\sgrav) and the next-to-LSP (NLSP) is
  the neutralino \(\snone\).  For further details please see some nice
  introductions to the subject 
  in~\cite{Martin:1997ns}. 
\par
\begin{figure}[h]
  \hspace{-0.15in}\includegraphics[width=0.5\columnwidth]{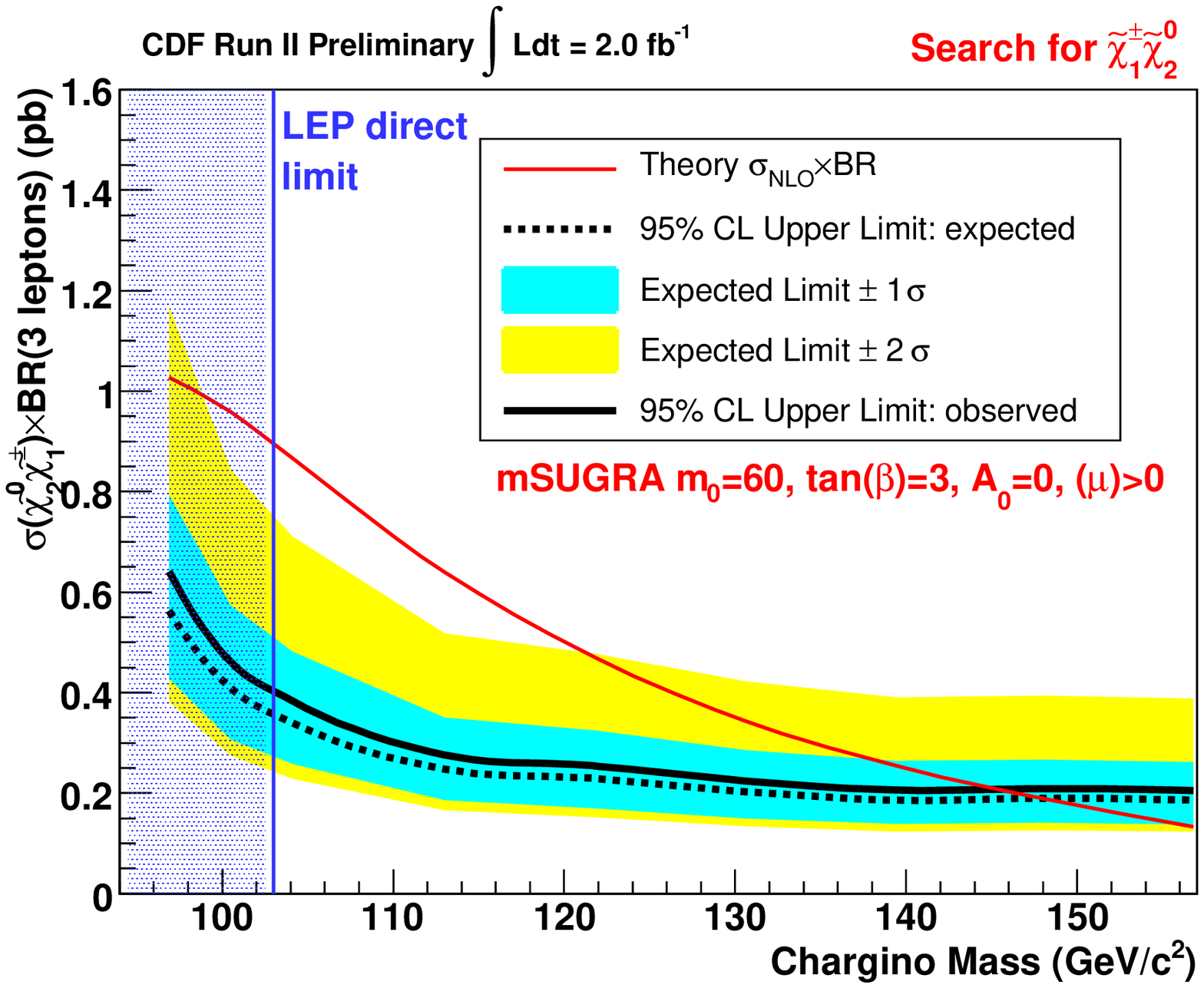}
  \hspace{-0.15in}\includegraphics[width=0.5\columnwidth]{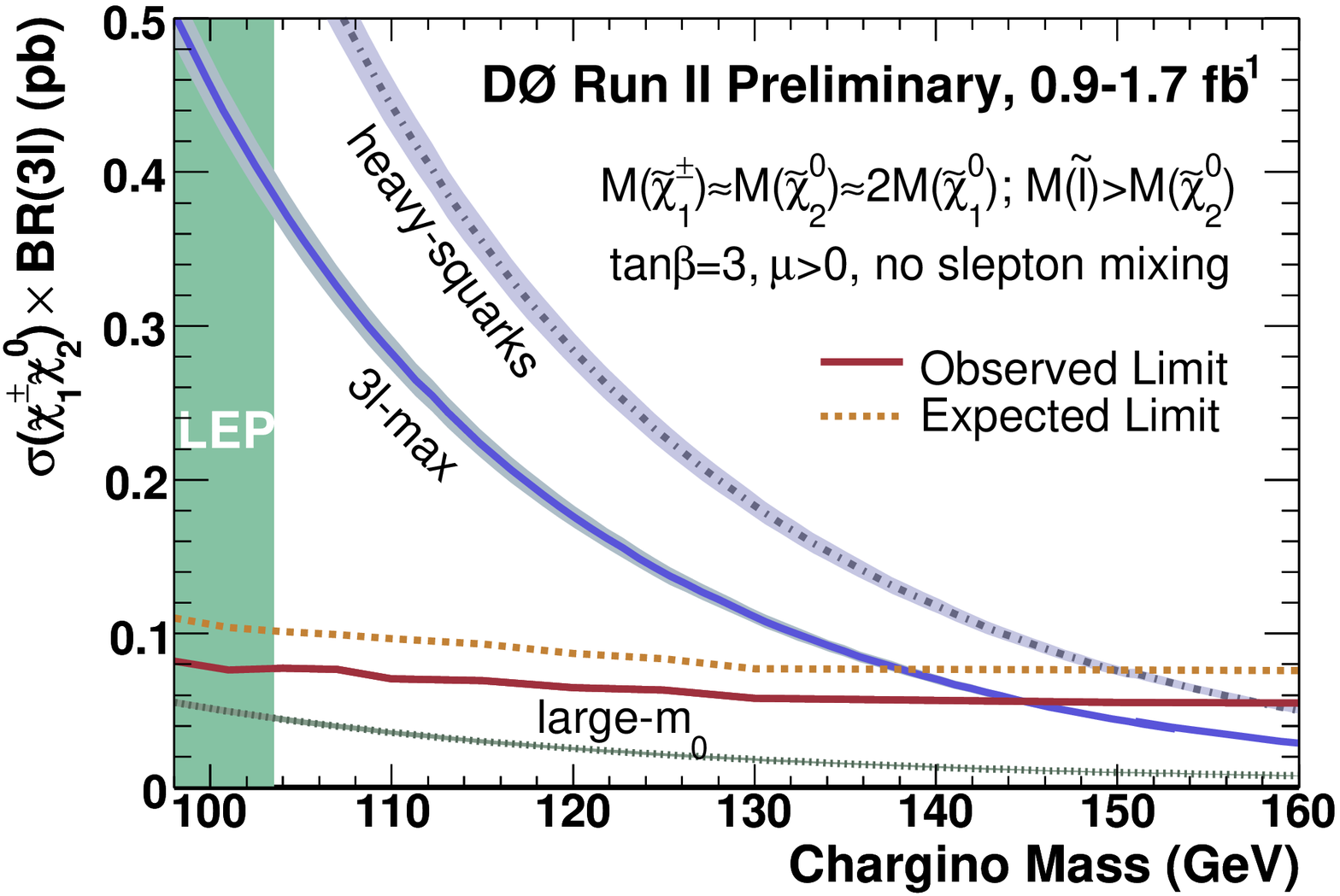}
  \caption{The search for chargino-neutralino production.
           CDF~\cite{cdf:3lepton} shows expected and observed limits
           for mSUGRA with \( m_{0}=60\gevcc \). CDF sets the lower
           limit for \( m(\schone)\,>145\gevcc \) at 95\% C.L.
           \dzero~\cite{d0:3lepton} uses several mSUGRA scenarios and
           for the \(3\lepton \)-max scenario 
           \(m(\schone)\,>140\gevcc\) at 95\% C.L. Please note the
           difference in mSUGRA benchmarks used by CDF and \dzero
           experiments.}
  \label{exp:3lepton}
  \end{figure}

  One of the promising production modes of SUSY is the pairs of 
  the lightest chargino \schone and next-to-lightest neutralino \sntwo
  decaying into three leptons and neutrinos with the LSP \snone 
  unobservable. The CDF analysis~\cite{cdf:3lepton} is based on  
  \( \sim2.0\invfb \) of integrated luminosity. The \dzero~Collaboration~\cite{d0:3lepton} 
  performed the trilepton searches with two samples of \(590\invpb \)
  and of \(1.0\invfb \) and combined the two results. Both experiments
  looked for modes in two categories -- channels with three leptons and 
  channels with two leptons and a single isolated track measured 
  by a tracking system. The leptons are \( \mmu \) identified by muon chambers 
  or \( \electron \) identified  as an electromagnetic cluster matched with 
  a track. Both analyses required large \( \etmiss\gsim\,20\gev\), high-\pt
  leptons, applied anti-\Z, anti-\ttbar, anti-\W cuts, and suppressed jet activity.
  Both experiments found in the selected signal regions good agreement
  between the number of observed events and 
  the event count predicted by the SM. The upper limits were set on the 
  production cross-sections in  the mSUGRA framework. Please
  see Fig.~\ref{exp:3lepton}.
\par
\begin{figure}[h]
  \hspace{-0.15in}\includegraphics[width=0.6\columnwidth]{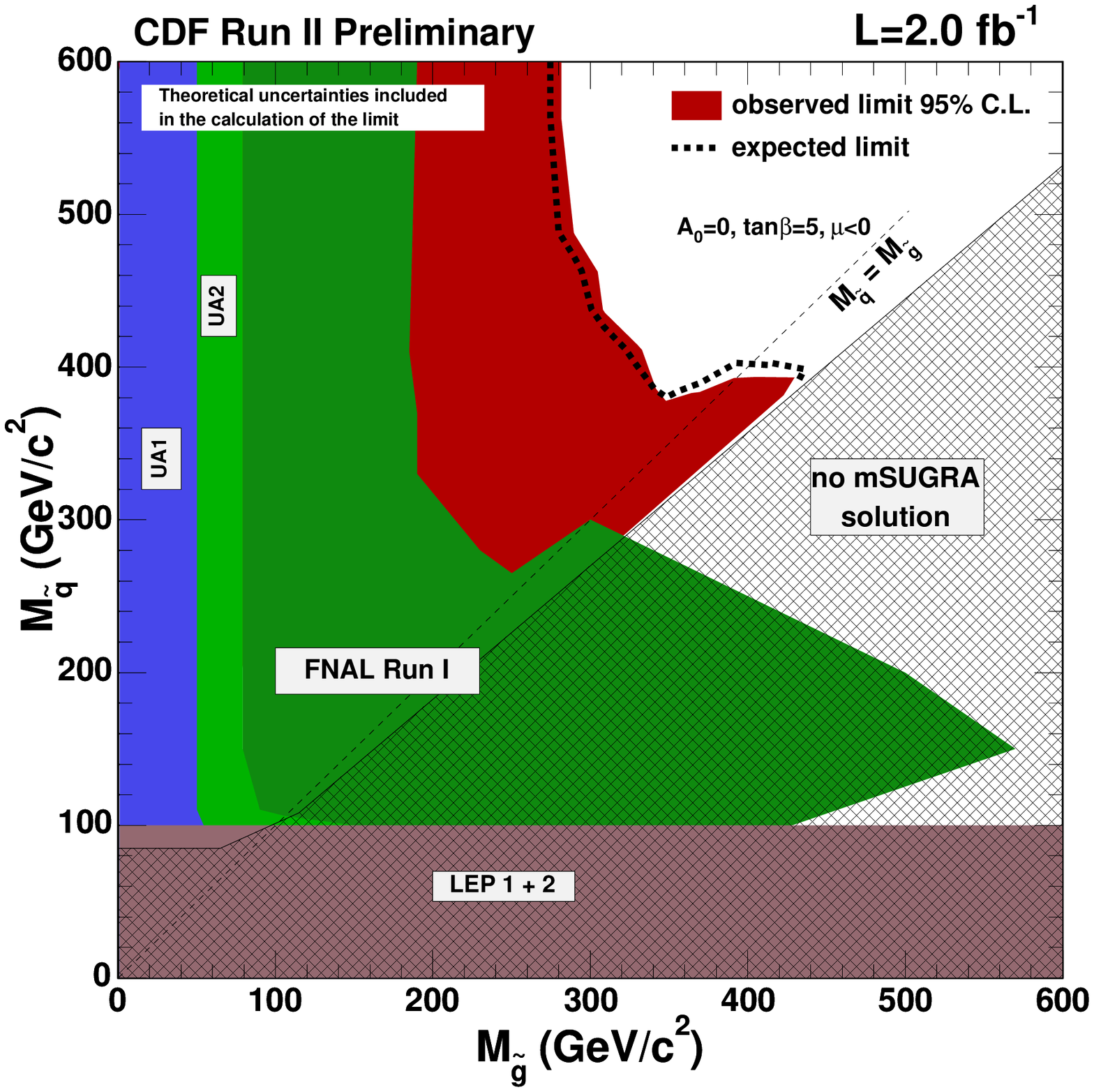}
  \hspace{-0.15in}\includegraphics[width=0.5\columnwidth]{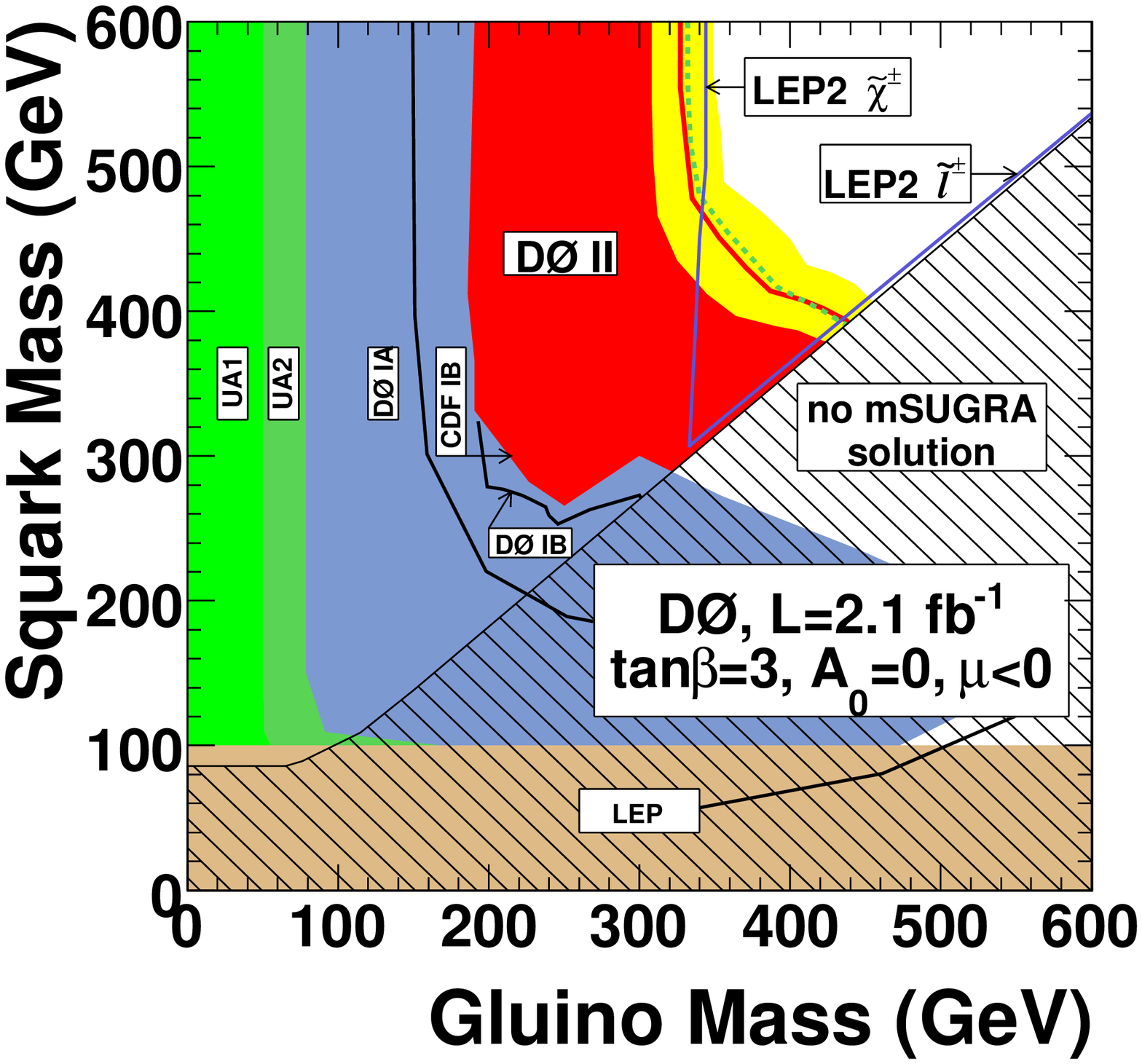}
  \caption{CDF and \dzero results: the observed (red) and expected (dashed line) exclusion 
           limits at the 95\% C.L. CDF uses mSUGRA with 
           \( \tan\beta=5,\,A_{0}=0,\,\mu_{0}<0 \) and sets lower limits~\cite{cdf:sq-glui}
           \( M > 392\gevcc\,@95\%\,\rm{C.L.} \) for \( M(\sglue)\simeq\,M(\sq) \),
           \( M(\sglue) > 280\gevcc\,@95\%\,\rm{C.L.} \) for every \( M(\sq) \) and
           \( M(\sglue) > 423\gevcc\,@95\%\,\rm{C.L.} \) for \( M(\sq)<378\gevcc \).
           \dzero~uses mSUGRA with \(\tan\beta=3,\,A_{0}=0,\,\mu_{0}<0 \) and sets 
           lower limits~\cite{:2007ww} \( M(\sq)>379\gevcc,\,M(\sglue)>308\gevcc\,@95\%\,\rm{C.L.} \)}
  \label{exp:sq-gluino}
  \end{figure}

  The \(\pap \) collisions at \(\sqs=1960\gev \) created by the Tevatron
  is a good place to search for the production of squarks (\sq) and
  gluinos (\sglue) in both CDF and \dzero detectors.  The optimism
  comes from the fact that the rates of \sq and \sglue are enhanced by
  a strong interaction \as involved in the processes. Depending on
  the mass relation between \sq and \sglue
  \((M(\sq)\,\lsim\mid\gsim\mid\,\simeq\,M(\sglue)) \) and availability
  of phase space to produce pairs of (\sq,\sq), (\sglue,\sglue) or
  (\sq,\sglue), their decays cascade to \(\geq\,2\)-, \(\geq\,4\)-
  or \(\geq\,3\)-~jet topologies with a large \etmiss caused by
  neutralinos \snone leaving no signals in the detectors.  CDF 
  analyzed~\cite{cdf:sq-glui} all three possible topologies using 
  \(\IntL=2\invfb \) of collected data. 
  The cuts to individual jet \( \et^{jet} \) optimized for every topology, 
  to missing \etmiss and total \Ht (\(\equiv\sum\et^{jet}\)) 
  have been applied together with
  lepton vetoes to suppress a SM contribution coming from \( \W\,/\,\Z\),
   \ttbar and QCD multi-jet events. The observed event count was
  found to be consistent with the SM background estimates and the
  exclusion limits at 95\%~C.L. were set in the \(M(\sq)\) vs
  \(M(\sglue)\) plane as shown in Fig.~\ref{exp:sq-gluino} together
  with a published analogous analysis from~\dzero~\cite{:2007ww}.
\par
\begin{figure}[h]
  \hspace{-0.15in}\includegraphics[width=0.5\columnwidth]{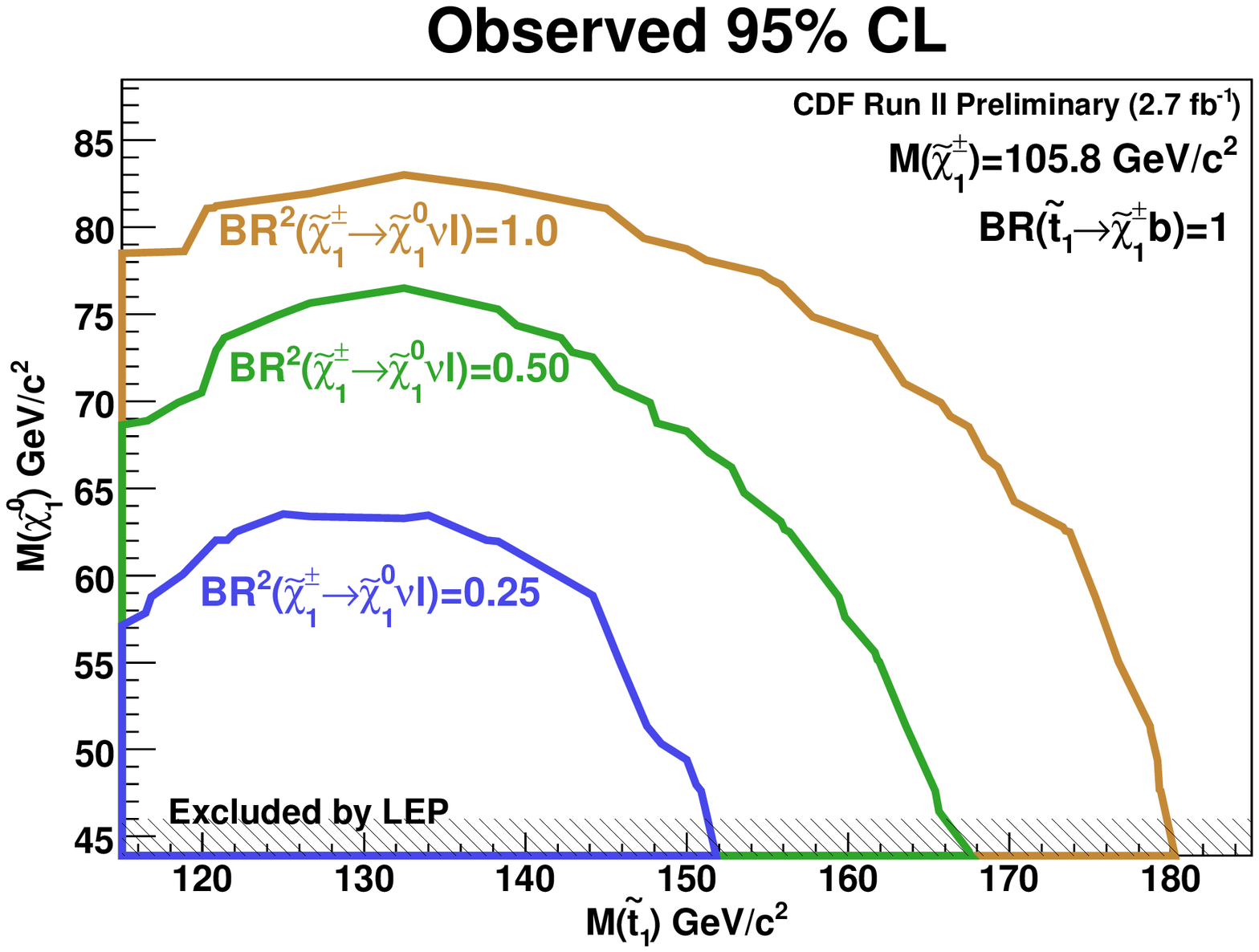}
  \hspace{-0.15in}\includegraphics[width=0.5\columnwidth]{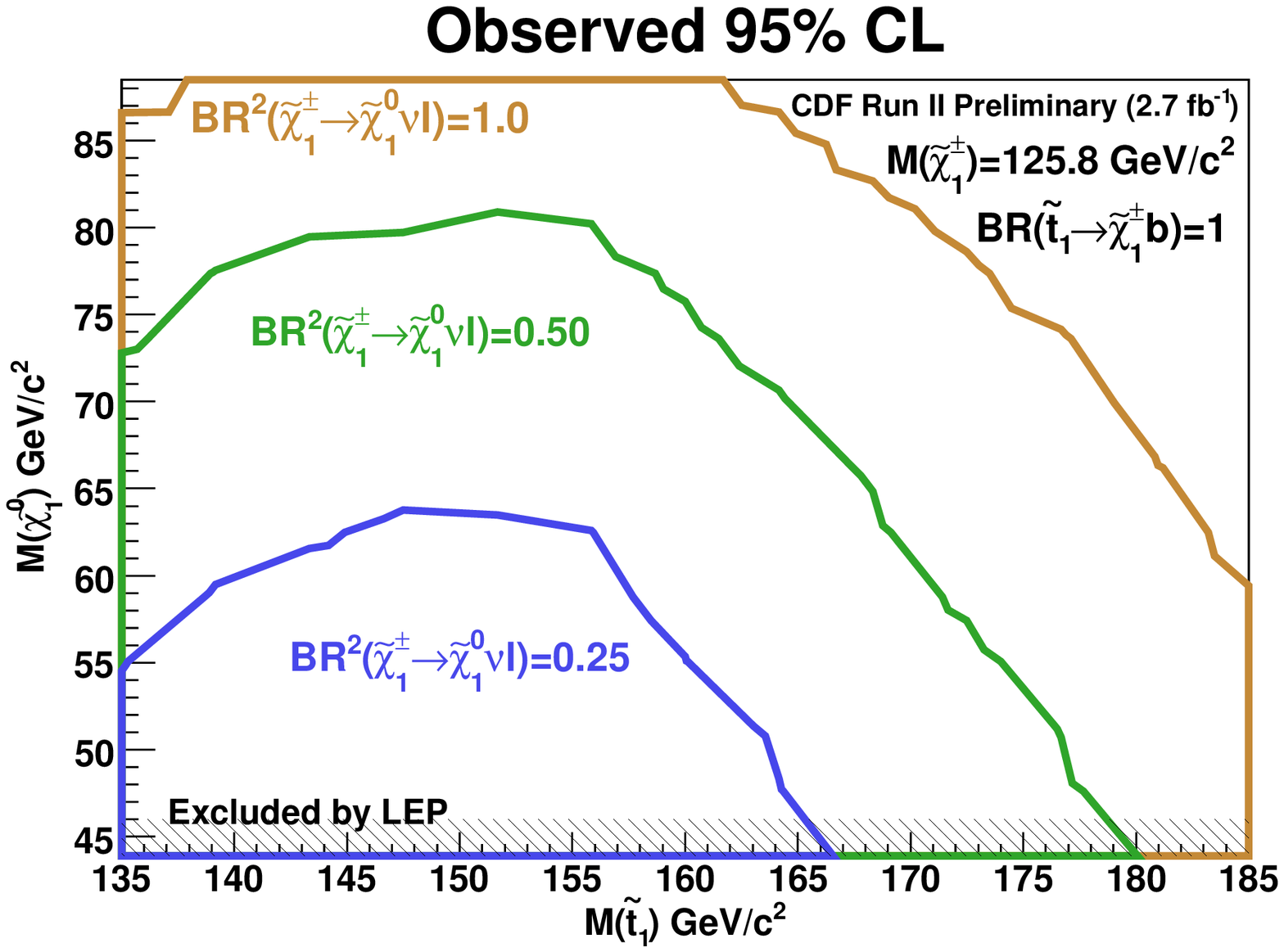}
  \caption{CDF preliminary results on the scalar top~\cite{cdf:stop-pair}. 
           M(\snone) vs M(\stopone) plane with regions excluded at 95\% C.L., 
            for various dilepton branching ratios, at \( M(\schone)=105.8\gev\) (left) 
             and at \( M(\schone)=125.8\gev\) (right). The electrons, muons and taus are 
            all assumed to equally contribute to the dilepton final state.}
  \label{cdf:stop-pairs}
  \end{figure}

  Within MSSM framework the SUSY partners of the \t-quark, scalar tops, are
  strongly mixed~\cite{Martin:1997ns} resulting in a significant splitting between
  eigenstates, \( M(\stopone) < M(\stoptwo) \). Moreover the \stopone
  could be the lightest \sq and even \( M(\stopone)\lsim\,m(\t) \).
  Based on data of \(\IntL=2.7\invfb\), CDF performed a search
  for the scalar top~\cite{cdf:stop-pair}.  The 2-body decay
  \(\stopone\to\b\schone \), is assumed to be dominant with
  \(\BR=100\% \), while \(\schone\to\snone\lepton\nul \) via a variety
  of modes. The assumption is valid provided \snone is LSP (mSUGRA),
   \(M(\stopone)\lsim\,m(\t)\) and \(M(\schone) < M(\stopone) - m(\b)\). 
  Then \( M(\stopone) \) is reconstructed kinematically in the
  assumed decay mode and the mass itself is used as a variable to
  discriminate stop from the SM background.  The experimental
  signature of produced pairs \(\stopone\overline{\stopone}\) is
  \(\lplm+{jet_{1}jet_{2}}+\etmiss \).  Two data samples, \b-tagged and
  anti-tagged, are considered with slightly different analysis
  cuts. Finally, the limits on the dilepton branching ratio at
  \(\sigma_{theor}(\stopone\overline{\stopone}) \) 
  for the 3-d mass area of
  \(M(\stopone)\in(115,185)\), \(M(\snone)\in(43.9,88.5)\) and 
  \(M(\schone)\in(105.8,125.8)\) are set. Please see
  Fig.~\ref{cdf:stop-pairs}. \dzero Collaboration performed a 
  similar analysis~\cite{d0:stop-pair} based on \(1.1\invfb \).
  The \stopone mode was assumed to be the same as CDF, but 
  the LSP in the SUSY benchmark was conjectured to be sneutrino (\snu). 
\par
  An interesting decay mode gets opened once we allow for the
  stop quark to decay as \(\stopone\to\c\snone \) via FCNC
  penguin diagram. \dzero Collaboration searched~\cite{Abazov:2008rc}
  for this mode assuming its \(\BR=100\%\) and analyzing the events
  with two charm acoplanar jets and \etmiss using a dataset of
  \(\IntL=1.0\invfb \).  With the theoretical
  uncertainty on the \(\sigma_{theor}(\stopone\overline{\stopone}) \)
  propagated properly, the largest limit was set for
  \( M(\stopone)>150\gev\) at 95\% C.L. for LSP mass set to 
  \( M(\snone)=65\gev \).
%
%
%
\par
\begin{figure}[htb]
  \hspace{-0.11in}\includegraphics[width=0.5\columnwidth]{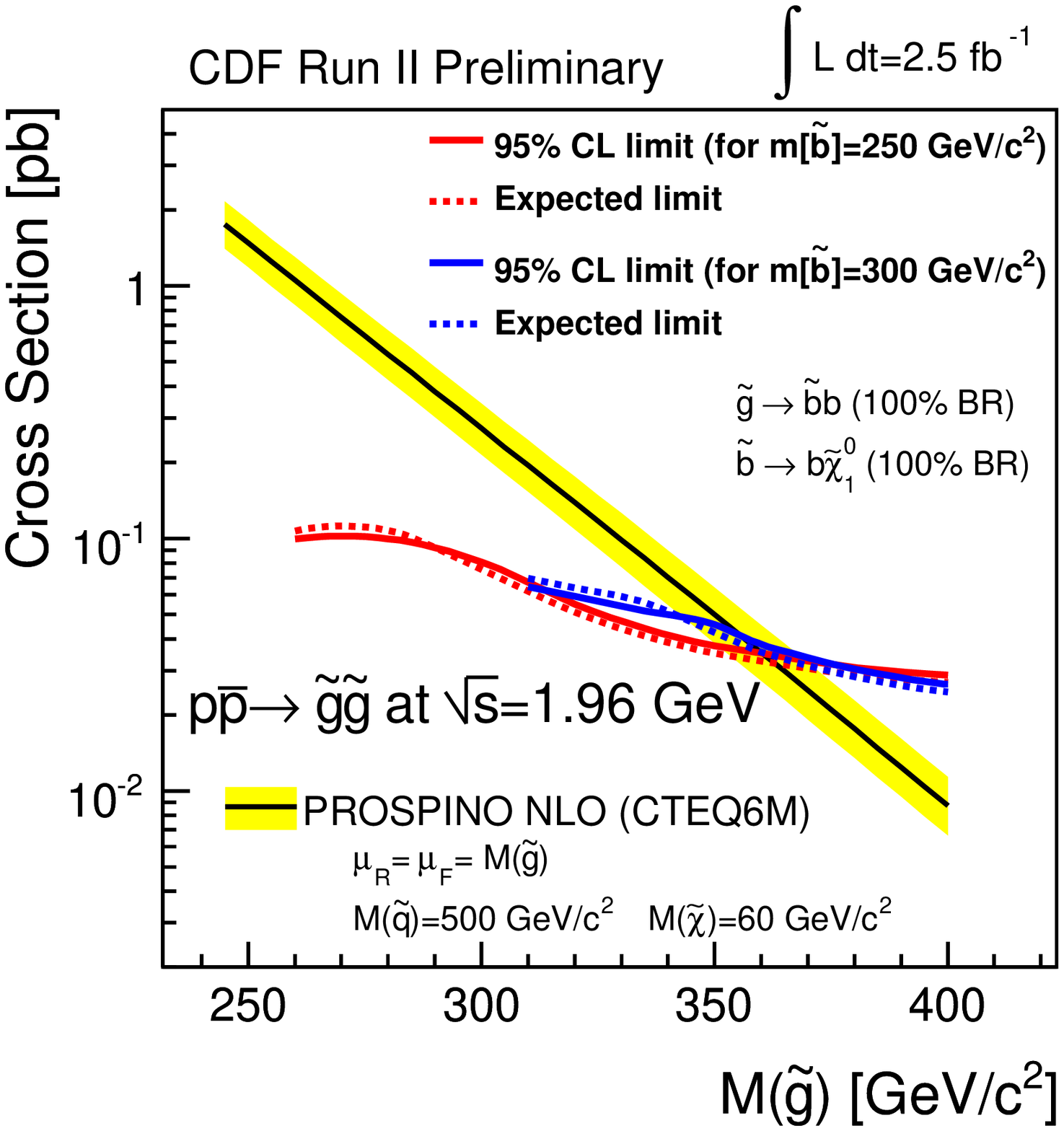}
  \hspace{-0.15in}\includegraphics[width=0.5\columnwidth]{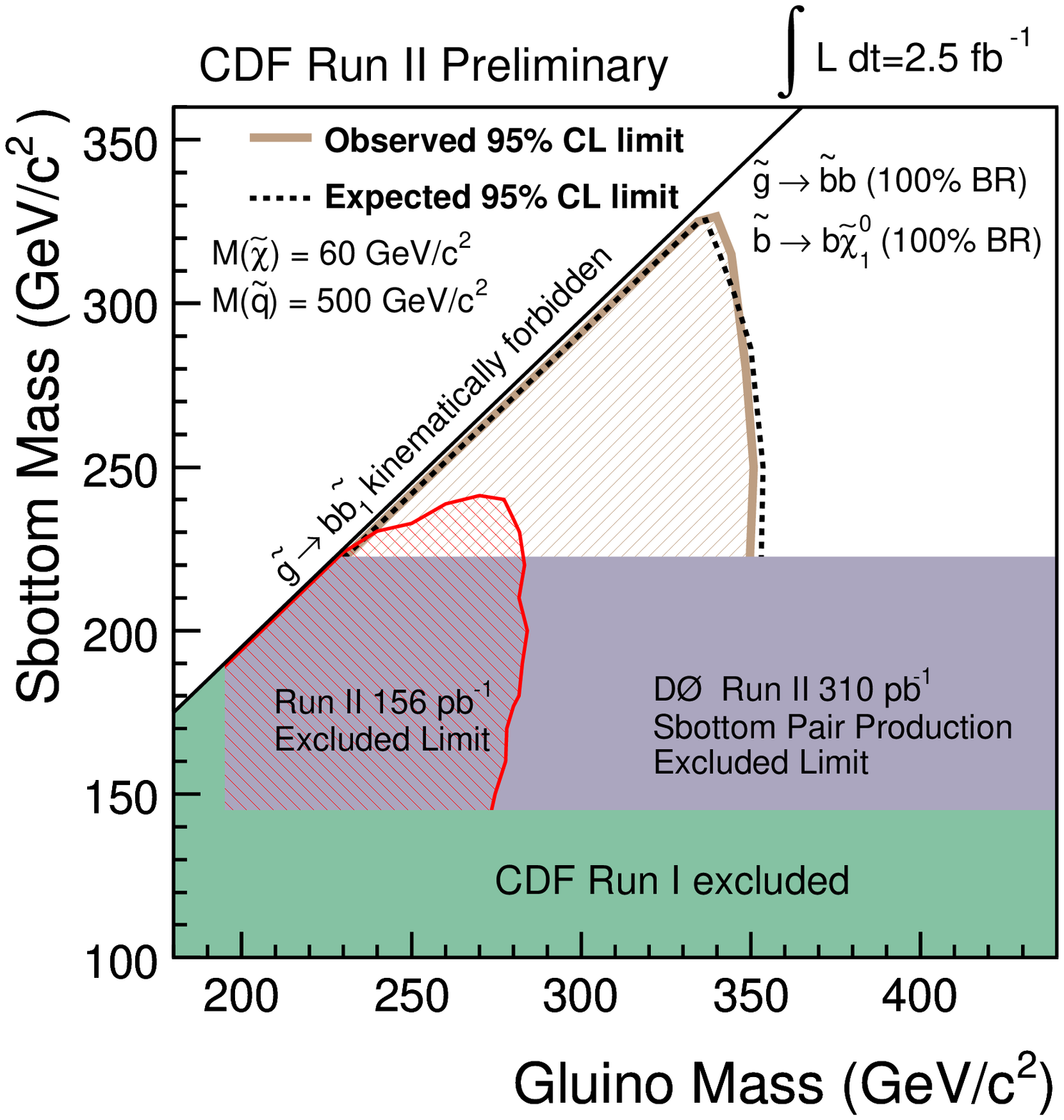}
  \caption{Left: Upper limit on \( \sigma(\,\pap\to\sglue\sglue,\,\sglue\to\b\sbotone\,(100\%)\,)\) at 95\% C.L.
           as a function of \(M(\sglue)\) for two different \(M({\sbotone})=250,300\gevcc \).
           Right: \(M({\sbotone})\,\rm{vs}\,M({\sglue}) \) exclusion plot showing observed and expected
           limits at 95\% C.L. with R-parity conserved. Please see the details in~\cite{cdf:glue-sbottom}. 
          } 
  \label{cdf:sbottom}
  \end{figure}

  The scalar bottom eigenstates predicted by MSSM are noticeably
  split given a strong mixing due to a large
  \(\tan{\beta}>10 \) and large negative contribution of corresponding
  top Yukawa coupling~\cite{Martin:1997ns}. 
  Consequently the lightest sbottom eigenstate \sbotone can be 
  reached at Tevatron energies. CDF used \(\IntL= 2.5\invfb\) of data collected by an
  inclusive \etmiss trigger to search~\cite{cdf:glue-sbottom} for the  \sbotone
  states {\it in decays of gluinos (\sglue)} expected to be abundantly produced at Tevatron,
  \( \pap\to\sglue\sglue\to\sbotone\overline{\sbotone} \).
  The \( M({\sglue})>M({\sbotone}) \) and further mass relations of 
  \( M({\stopone}),\,M({\schone})\,>M({\sbotone})\,>M({\snone}) \) allow for
  the decay mode with \(\BR(\sglue\to\b\sbotone,\,\sbotone\to\b\snone)=100\% \)
  to be dominant. This area of phase space was probed by the data analysis.
  The experimental signature of the pair \(\sbotone\overline{\sbotone} \) 
  is {{4\b-jets}\(+\etmiss\)} 
  and CDF required at least 2~\b-jets to be \b-tags. No significant deviation 
  from the SM was observed and the exclusion limits were set using only the 
  2~\b-tags sample. Please see Fig.~\ref{cdf:sbottom}.
\par
\begin{figure}[htb]
  \hspace{-0.15in}\includegraphics[width=0.5\columnwidth]{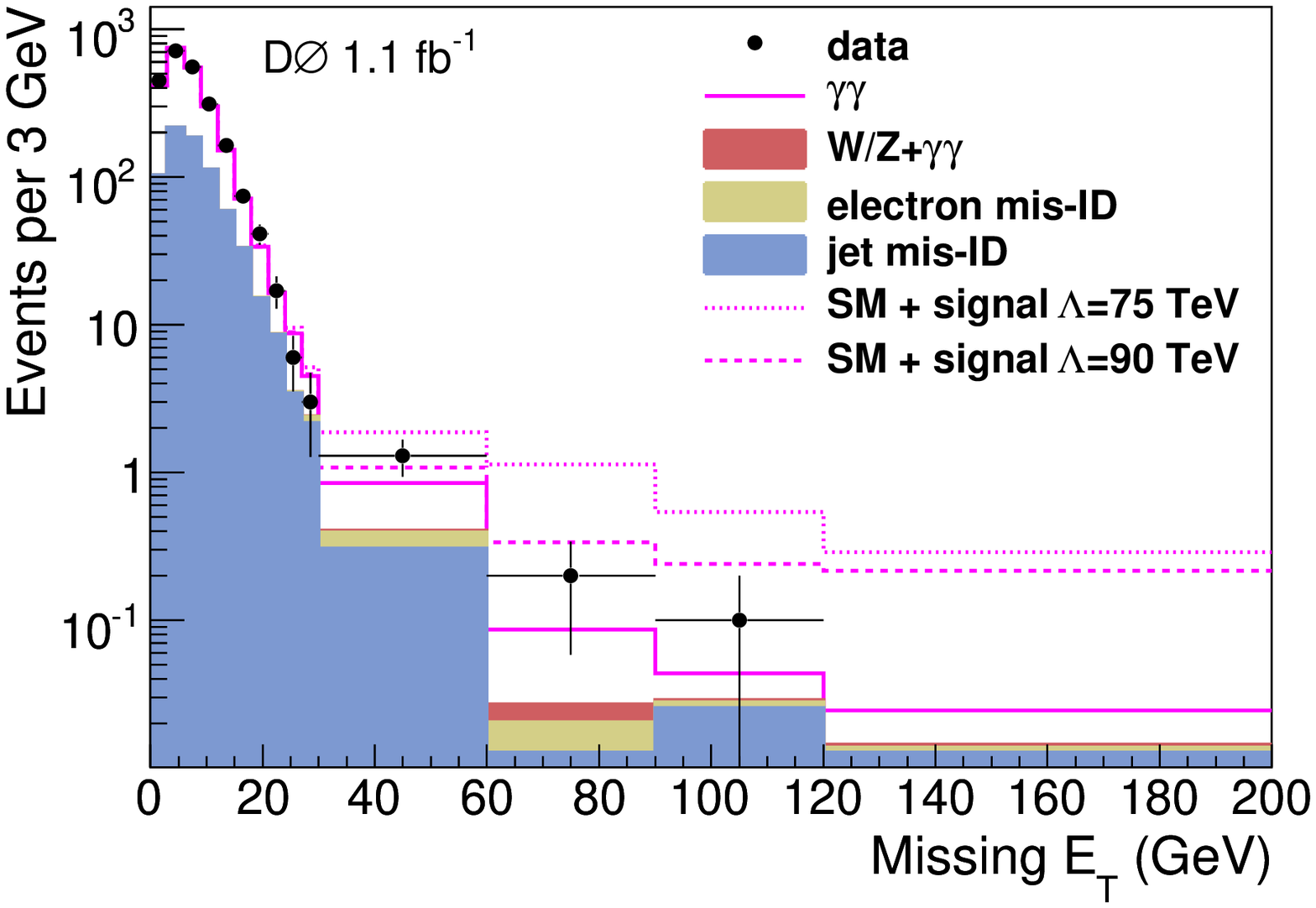}
  \hspace{-0.11in}\includegraphics[width=0.5\columnwidth]{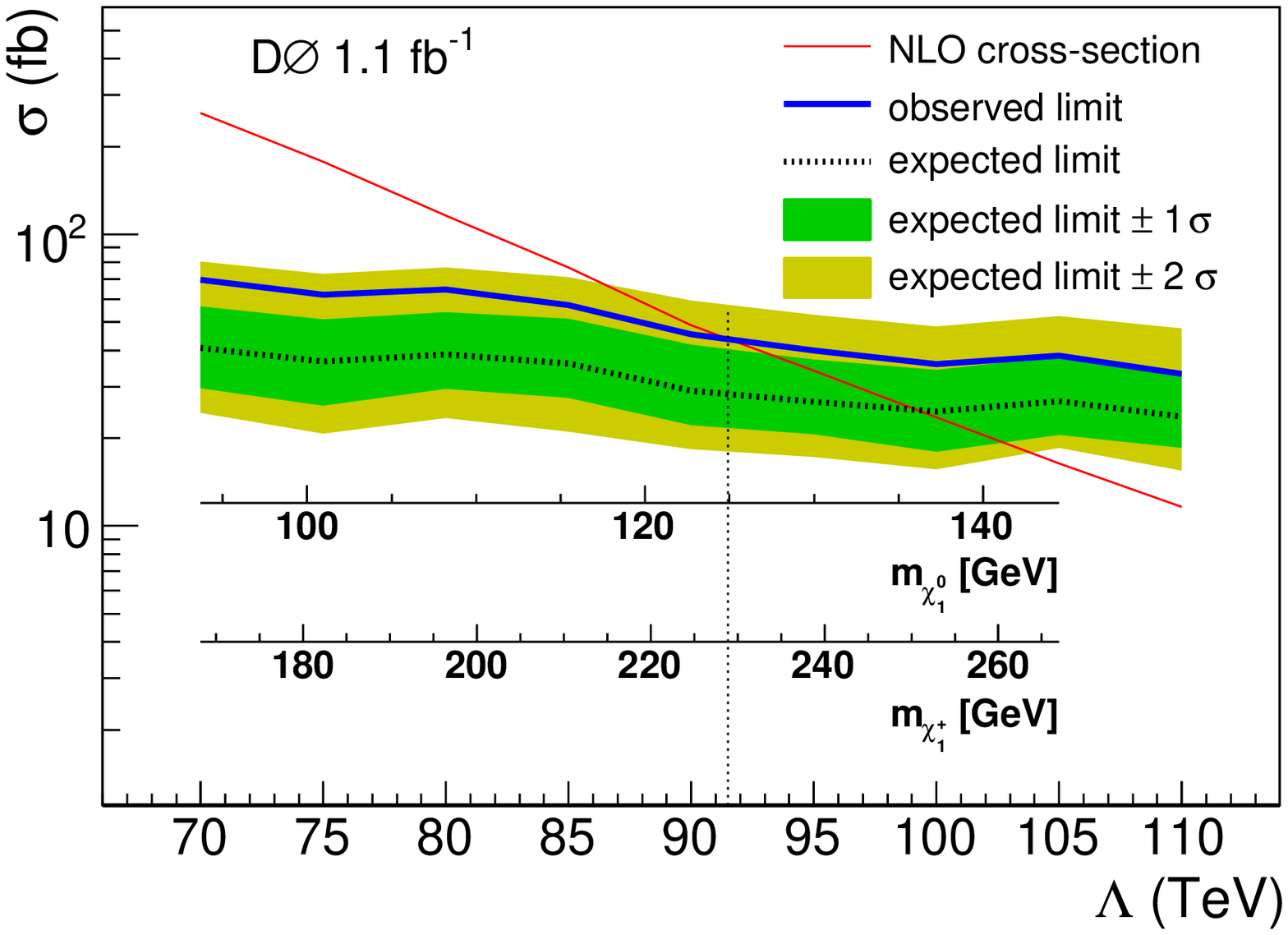}
  \caption{Left: The \( \etmiss \) distribution in \(\gaga\) data together with various SM background 
           contributions. The dotted-dashed line shows the GMSB SUSY theoretical prediction.
          Right: The predicted cross section for the Snowmass Slope model versus \( \Lambda \)
          (see~\cite{:2007is} and references therein). The observed (solid line) 
          and expected (dash-dotted) upper limits at \( 95\% \) C.L. are shown.
          Please see the details in~\cite{:2007is}. } 
  \label{d0:gmsb}
  \end{figure}

  The analyses presented above used an mSUGRA as a benchmark to set
  exclusion areas on mass plots or limits on cross-sections.  \dzero
  Collaboration has undertaken a search~\cite{:2007is} for 
  signatures of MSSM with GMSB breaking scenario. LSP in GMSB is the
  gravitino \sgrav. A neutralino (NLSP) having admixture of photino
  decays predominantly in the mode \(\snone\to\g\sgrav \).  Hence,
  assuming \(R_{P}\) invariance, the GMSB event should have a signature
  of \( \gaga + \etmiss \) in a final state. \dzero Collaboration has
  analyzed \(\IntL=1.1\invfb \) of data. Two photons above 25\gev have
  been selected and an excess of events over SM backgrounds was
  searched in \(\etmiss >60\gev \) range. Please see the left plot at
  Fig.~\ref{d0:gmsb}. No significant excess of the event count in
  \etmiss spectrum over expected SM background was found. The most
  stringent lower limits on the GMSB signal cross section to date were
  set for gaugino masses \(M(\snone)\gsim125\gev \) and
  \(M(\schone)\gsim229\gev \). See the right plot at
  Fig.~\ref{d0:gmsb}.
\section{Non-SUSY Searches at the Tevatron}
  One of the exciting topics among non-SUSY searches is the one
  concerned with Large Extra Dimensions (or LED). A theory 
  of LED has been outlined by the 
  authors~\cite{ArkaniHamed:1998rs}.
  From an experimental point of view the basic subprocesses 
  where we could expect the LED to reveal itself
  are the mono-photons in \( { q\stilde{q}\to\g\,\KKgrav} \) or
  mono-jets in \( { q\glue\to\q\,\KKgrav} \),
   \( {\glue\glue\to\glue\,\KKgrav} \) always associated with large
  \etmiss.  \dzero searched for events with \( \g +\etmiss \) in a
  final state~\cite{d0:led-2-7fb} using the data
  of \(\IntL=2.7\invfb \) collected with a trigger on
  electromagnetic clusters of \( \et(\g) >20\gev \). The analyzed
  photon sample consisted of events with only a single photon having
  transverse momentum \( \pt (\g) >90\gev \) and with \( \etmiss >70\gev \).
  The CDF Collaboration in a similar analysis~\cite{cdf:led} based on 2.0\invfb
  combined both single \g and mono-jet results and set the limits.
  The results of both analyses are summarized in Table~\ref{exp:led}.
  \begin{table}[h]
    \begin{center} 
    \begin{tabular}{|c|c|c|c|c|} 
    \hline
    \multicolumn{1}{|c|}{Analysis:} & \multicolumn{2}{c|}{\( \dzero,\,{\g +\etmiss},\,2.7\invfb \)} & 
    \multicolumn{2}{c|}{CDF Prelim., Jet/\( \mathbf{\g +\etmiss},\,2.0\invfb \)} \\ 
    \hline\hline
    $\mathbf{N_{LED}}$ & $\sigma_{obs(exp)}^{95}$ & $M_D^{obs(exp)}$ & $\sigma_{obs(exp)}^{95}$ & $M_D^{obs(exp)}$ \\
    \hline
    $2$&$19.0~(14.6)$&$970~(1037)$&$26.3$ & $1420$\\
    $3$&$20.1~(14.7)$&$899~(957)$&$38.7$ & $1160$\\
    $4$&$20.1~(14.9)$&$867~(916)$&$46.9$ & $1060$\\
    $5$&$19.9~(15.0)$&$848~(883)$&$52.7$ & $990$\\
    $6$&$18.2~(15.2)$&$831~(850)$&$56.7$ & $950$\\
    $7$&$15.9~(14.9)$&$834~(841)$&       &      \\
    $8$&$17.3~(15.0)$&$804~(816)$&       &      \\
    \hline
    \end{tabular}
    \end{center}
    \caption{The upper limit at 95\% C.L. set for a cross-section 
             by \dzero~\cite{d0:led-2-7fb} and CDF~\cite{cdf:led} 
             and the  95\% C.L. lower limit set for a fundamental 
             Planck scale \( M_{D} \) are shown in the table.}
    \label{exp:led} 
  \end{table}
\par
  Exotic models like \(\mathrm{GUT\,E_{6}} \) extension of Standard
  Model or quantum gravitational Randall-Sundrum model predict high
  mass resonances in di-lepton modes 
  \(  \Zprim\to\lepton^{+}\lepton^{-} \)~\cite{Hewett:1988xc}.
  Using a total luminosity \( \IntL=2.5\invfb \), CDF
  searched for dielectron resonance candidates of \( \Zprim\to\epem \)~\cite{cdf:zprim-ee}.
  Given that the search probes the mass range \( M(\epem)\in\,(150,1000)\gevcc \) 
  an excess of \( \sim3.8\sigma \) was found in the \( M(\epem)\sim240\gevcc \) 
  region with  \( \alpha=0.6\% \) caused by the background fluctuation.
  The lower mass limit on SM coupling \Zprim, \( M(\Zprim_{SM})>966\gevcc \) 
  was set at 95\% C.L. The RS graviton with a  mass below 850\gevcc 
  was excluded at 95\% C.L. assuming \( k/{{\overline{M}}_{Planck}}=0.1 \). 
%
%
%
%
%
\section*{Acknowledgements} 
  The author is grateful to the members of CDF and \dzero beyond
  Standard Model working groups for their useful suggestions and
  comments made during the preparation of this talk. The author thanks
  J.~E.~Metcalfe (Univ. of New Mexico) for her comments to the
  write-up of the talk.  The author is thankful to S.~C.~Seidel (CDF
  Collab.)  for supporting this work.
%
%
%

%

\begin{thebibliography}{999}
\bibitem{def:Rp}
\( R_{P} = (-1)^{3(B-L)+2S} \) for a particle with spin \( S \).
Following \( B-L \) invariance, \( R_{P} \) is conserved in MSSM. \( R_{P}=+1 \) for SM particle,
and \( R_{P}=-1 \) for its SUSY partner.
\end{thebibliography}
\end{document}